\documentstyle[prl,twocolumn,aps,color]{revtex}
\definecolor{rosso}{rgb}{1,0,0}
\definecolor{orang}{rgb}{.48,.38,.14}
\definecolor{ciano}{rgb}{0,1,1}
\definecolor{bluuu}{rgb}{0,0,1}
\definecolor{viola}{rgb}{1,0,1}
\input epsf
\begin{document}
\draft
\twocolumn[\hsize\textwidth\columnwidth\hsize\csname@twocolumnfalse\endcsname

\title{Comparative genomics study of inverted repeats in bacteria}
 
\author{Fabrizio Lillo $^{*}$, Salvatore Basile $^{*,\dag}$ and Rosario N.
Mantegna $^{*,\dag}$ }

\address{
$^{*}$Istituto Nazionale per la Fisica della Materia, Unit\`a di Palermo, 
Viale delle Scienze, I-90128, Palermo, Italia\\ 
$^{\dag}$ Dipartimento di Fisica e Tecnologie Relative, 
Universit\`a di Palermo, Viale delle Scienze, I-90128
Palermo, Italia}

 
\maketitle
 
\begin{abstract} We investigate the number of inverted
repeats observed in 37 complete genomes of bacteria.
The number of inverted repeats
observed is much higher 
than expected using Markovian models
of DNA sequences in most of the eubacteria. 
By using the information annotated 
in the genomes we discover that in most of the 
eubacteria the inverted repeats of stem length
longer than 8 nucleotides preferentially 
locate near the 3' end of the nearest coding regions.
We also show that IRs characterized by 
large values of the stem length locate preferentially
in short non-coding regions bounded by two 3' ends 
of convergent genes. 
By using the program TransTerm recently introduced to 
predict transcription terminators in bacterial genomes,
we conclude that only a part of the observed 
inverted repeats fullfills the model requirements 
characterizing rho-independent termination in 
several genomes.
\end{abstract} 

\vskip2pc]

An inverted repeat (IR) in DNA sequences  
provide the necessary condition for the potential existence
of a hairpin structure in the transcribed messenger 
RNA and/or cruciform structures in DNA 
\cite{Sinden94}. Inverted repeats play an 
important role for regulation of transcription 
and translation. Examples are the role of the IR
located in the promoter region of the S10 ribosomal
protein operon \cite{Post78,Shen88,Li97} and the role
of IR in the $lac$ operator \cite{Sadler83,Simons84}. 
It has also been proposed that hairpins play an important 
role in the control of transcription, translation and
other biological functions \cite{Raghunathan91,Blatt93}.
Hairpin- or cruciform-binding proteins have been identified from
several species. These results suggest that regulatory hairpins
may be involved in transcription of a number of genes
\cite{Wadkins2000}. 
The existence in vivo
of hairpin structures of messenger RNA during transcription 
has been demonstrated 
\cite{Farnaham81,Platt86,Yager91,Kroll92,Wilson95}. 
Hairpin structures are often associated with 
rho-independent intrinsic 
terminators of genes in several bacteria.
The presence of such 
intrinsic terminators has been observed in {\it E. coli}
\cite{Carafa90}.
Rho-independent intrinsic terminators have also been detected in other 
bacteria as, for example, 
{\it Streptococcus pneumoniae} \cite{Diaz90}, 
{\it Pseudomonas aeruginosa} \cite{Gamper92}, 
{\it Myxococcus xanthus} \cite{Kimsey92},
{\it Streptococcus equisimilis} H46A \cite{Steiner97} and 
{\it Chromatium vinosum} D \cite{Pattara98}.
Hairpin structures can occur at the mRNA when an IR 
is present in the DNA sequence. For example, the sequence 
5'aGGAATCGATCTTaacgAAGATCGATTCCa3' is a sequence having a sub-sequence
GGAATCGATCTT which is the IR of AAGATCGATTCC. This 
IR can form a hairpin having a stem of length 
12 nucleotides and a loop (aacg) of length 4  nucleotide 
in the transcribed RNA. The number of IRs
has been investigated with bioinformatics methods  
in long DNA sequences of eukaryotic (human and yeast) and bacterial 
({\it E.coli}) DNA \cite{Schroth95} and in the complete genomes of
eubacterium {\it Haemophilus influenzae} \cite{Smith95}, archaebacterium
{\it Methanococcus jannaschii} and cyanobacterium
{\it Synechocystis} sp. PCC6803 \cite{Cox97}. These studies
have shown that inverted repeats are rather abundant 
in {\it E. coli} and {\it Haemophilus influenzae}, poorly abundant in 
{\it Methanococcus jannaschii} and with no enrichment
(with respect to a Bernoullian assumption about DNA sequences) 
in {\it Synechocystis} \cite{Schroth95,Cox97}.

\section*{Data and Methods}

In the present study, we investigate 37 complete genomes of bacteria
recently sequenced (all the bacterial complete genomes
publicly available at the time of our study). The set consists 
of 8 archaebacteria, 
1 aquificales, 1 thermotogales, 2 spirochetales, 5 chlamydiales,
1 deinococcus, 1 cyanobacterium, 12 proteobacteria (purple bacteria)
and 6 firmicutes (gram positive). The analyzed DNA totals 77.8 millions 
base pairs. In the completed genomes we search with a specialized
computer program all the inverted repeats of stem
length $\ell$ ranging from 4 to 20 and loop (spacer) length $m$ ranging
from 3 to 10. These are typical boundaries in the range of 
the ones used in the literature for the investigation of IRs,
in DNA sequences \cite{Schroth95,Cox97,Ermolaeva2000}.
The results of our investigation are illustrated in Figure 1
and summarized in Table I. 
The available
genomes differ the one from the other with respect to the 
CG content and the degree of its fluctuation along the genome.
In our study, each detected IR is the IR
of maximal stem length $\ell$ located in a given DNA
position. In other words, we check that the first two nucleotides 
immediately out 
of the stem region have not a palindromic counterpart 
for each observed IR. 
Within this definition, the number of IRs expected in a 
genome under the simplest assumption of a random 
Bernoullian DNA is given by the 
equation     
\begin{eqnarray}
n_{ex}(\ell,m) & = & N (1-2P_aP_t-2P_cP_g)^2 \times \nonumber \\
& \times &  (2P_aP_t+2P_cP_g)^{\ell},
\end{eqnarray}
where $N$ is the number of nucleotides in the genome sequence 
and $P_a$, $P_c$, $P_g$ and $P_t$ are the observed 
frequencies of nucleotides.
Eq. (1) shows that the number of expected IRs 
is independent of $m$ whereas it depends on the CG content
of the genome.
The CG content can vary considerably across different 
regions of the same genome. Moreover, a Bernoullian 
description of DNA sequences provides just a 
a rough approximation of the statistical 
properties observed in real genome. For this reason, we 
decide to compare the results obtained in real genomes 
with the ones obtained by generating a computer generated first-order
Markov genome having the same probability matrix of 
dinucleotides empirically observed within each 
non-overlapping window of 
10,000 nucleotides for each genome. The occurrence of 
IRs detected in the computer generated genomes 
are used for comparison in Figure 1 and to obtain the 
$\chi^2$ values used to illustrate the empirical results
summarized in Table I.
\begin{figure}[t]
\epsfxsize=3in
\epsfbox{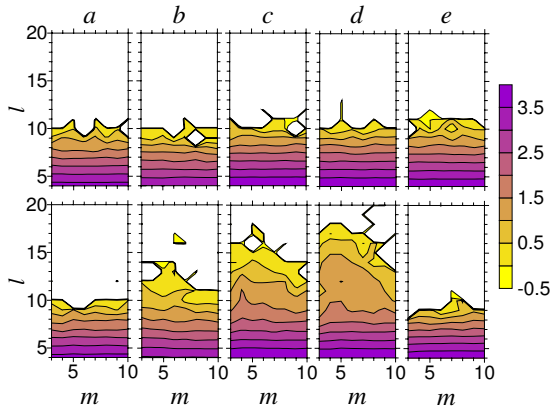}
\vspace{0.3cm}
\caption{Contour plot of the decimal logarithm of the total occurrence 
of IRs of stem length $\ell$ and loop length $m$ for 5 representative genomes 
(bottom panels) compared with the decimal logarithmic occurrence 
of corresponding computer generated first-order Markovian data 
(top panels). From left to right we show (a) the archaebacterium
{\it Archaeoglobus fulgidus}, (b) the {\it Chlamydia pneumoniae} CWL029,
(c) the proteobacterium {\it Escherichia coli}, (d) the firmicute 
{\it Bacillus subtilis} and (e) the cyanobacterium {\it Synechocystis} sp..
The numbers of color scale are decimal logarithm of the total 
occurrence of IRs.}
\label{fig1}
\end{figure}
In Figure 1 we show the contour lines of the decimal logarithm
of the number of occurrence of IRs 
as a function of the stem length $\ell$
and of the loop length $m$ for 5 representative genomes. 
To take into account the differences in length, CG 
content and Markovian 
characterization observed in the genomes we also show 
in Figure 1 (top panels of the figure) the results 
obtained for the same investigation performed with 
the corresponding computer generated Markovian genomes.
The selected genomes are: {\it Archaeoglobus fulgidus} 
(archaebacteria), {\it Chlamydia pneumoniae} CWL029 (chlamydiales), 
{\it Bacillus subtilis} (firmicutes), 
{\it Escherichia coli} (proteobacteria)
and {\it Synechocystis} sp. (cyanobacteriae).
In the absence of the statistical uncertainties detected 
at the noise level, Markovian generated data show
contour lines, which are straight lines parallel 
to the horizontal axis (loop length $m$). This is 
consistent with the theoretical result
of Eq. (1) obtained for a Bernoullian DNA sequence.
In fact, the theoretical prediction of Eq. (1)
does not depends on $m$. 
Moreover, the distance between two successive 
contour lines is approximately constant. This 
second observation indicates 
that, once again consistently with the prediction of Eq. (1),
the number of occurrences of inverted repeats
exponentially decreases with the stem length $\ell$ 
in a local Markov model of DNA sequences.     
The results obtained in real genomes (bottom panels
of Figure 1) are genome dependent. Some genomes as  
{\it Archaeoglobus fulgidus} and {\it Synechocystis} sp. 
show a general pattern hardly distinguishable from the 
one observed in computer generated data
whereas {\it Chlamydia pneumoniae} CWL029, 
{\it Bacillus subtilis} and 
{\it Escherichia coli} are characterized 
by the occurrence of inverted repeats which are not
explained by a first-order Markovian model
of DNA. For these three genomes, but they are 
representative of several others genomes, it is 
worth noting that the contour lines of Figure 1 show both
(i) a value of the occurrence of inverted repeats much 
larger than expected for a first-order 
Markovian model for large values of $\ell$ and 
(ii) an occurrence of inverted repeats 
which is strongly dependent on the specific value of
$m$ for large values of $\ell$.
We verify that higher-order Markovian models up to the
fifth-order also fail to reproduce these empirical results.
In other words these empirical results cannot be
ascribed to the strong bias up to the esamer level
present in several genomes. 
In agreement with previous studies devoted to the 
search of inverted repeats in long DNA sequences
(Schroth and Shing Ho, 1995; Cox and Mirkin, 1997) 
we detect the existence of
different levels of enhancement of the number of
observed inverted repeats in different species
of bacteria.  
The results presented in Figure 1 are just illustrative
of the varied behavior observed in 5 different case.
The complete results obtained by our investigation are
summarized in Table I where we group the 37 investigated 
genomes.

\section*{Results and Discussion}

In Table I we report the observed number of inverted
repeats as a function of the stem length $\ell$ for each
investigated genome. The genomes are listed in groups 
separated by a horizontal line. From top to bottom, 
the groups are archaebacteria, chlamydiales, firmicutes, 
proteobacteria and others. 
This last group includes aquificales, spirochetales,
deinococcales, cyanobacteria and thermotogales. 
To limit the size of the Table
we sum up the occurrence observed for different 
values of $3\le m \le 10$. To make comparison possible 
between genomes of different sizes, the values are 
normalized to one million base pairs. In the Table we also provide 
(in parenthesis) the value expected for computer generated
genomes characterized by a local first-order 
Markovian model. 
The $\chi^2$ values are calculated 
by comparing the empirical occurrence of IR with respect 
to the one predicted by a first-order Markovian process.
 The $\chi^2$ is 
calculated by comparing the empirical occurrence of 
inverted repeats with the occurrence of computer generated data
when $\ell$ is varying from 4 to 20 (the number of IRs
with $\ell=4$ and $\ell=5$ are not shown in Table I 
for lack of space. They are available at the web site
http://lagash.dft.unipa.it/IR.html). The $\chi^2$ calculation
is done by 
comparing the two distributions using six 
bins. In fact, the 
number of IRs with $\ell>8$ are summed up together in the
presentation of Table I and in the $\chi^2$
calculation to compensate the 
exponential decrease observed in the number of IRs
when $\ell$ increases .  
The obtained $\chi^2$ values are in all 
except one case larger than 30 implying that the 
$p$ value is always below (and often much below)
$1 \times 10^{-5}$. The only exception is the
aquificales {\it Aquifex aeolicus} that present a $p$ value
of 0.52. 
To make a direct comparison between 
different genomes, in Table I we use a color code 
for the contribution to the $\chi^2$ of each different 
number of IRs of stem length $\ell$. Specifically, for 
each value of $\ell$ we compute  
$(n_{obs}-n_{Markov})/\sqrt{n_{Markov}}$. The square of 
this quantity directly contributes to the $\chi^2$ value.
In the Table, for values of this parameter lower than
-3 we use a blue character. A black character is used 
when this parameter is between -3 and 3, for values
larger than 3 we use a red character and for values
larger than 10 the number of IRs is purple.  

Our results show that the number of 
inverted repeats is much higher than 
the one theoretically expected
for a first-order Markovian DNA of the same composition in 
chlamydiales, firmicutes and proteobacteria.
Deviations in the observed number of IRs that contribute
more than 100 to the $\chi^2$ are detected in several genomes
especially when $\ell>8$. Moreover,  we observe deviations 
of the number of IRs that contribute more than
9 to the $\chi^2$ in the large majority of 
genomes for almost all the $\ell$ values.
Among eubacteria
only {\it Aquifex aeolicus} show no enrichment, whereas
{\it Deinococcus radiodurans} 
and {\it Synechocystis} sp. show less IRs than predicted by 
a first-order Markovian DNA.
In contrast, a general pattern does not emerges in 
archaebacteria.
In three of the eight complete genomes investigated 
the number of empirically 
observed IRs is comparable with the one expected by using a 
Markovian model, whereas in the cases of 
{\it Aeropyrum pernix}, {\it Halobacterium sp.}, 
{\it Methanococcus jannaschii},
{\it Methanobacterium thermoautotrophicum}
and {\it Thermoplasma acidophilum} IRs are slightly more frequent
or more frequent  
than expected. In spite of this variety of behavior, a general
remark is that
the number of IRs in archaebacteria is, in most cases,
much lower than the one observed in eubacteria.
\begin{figure}[t]
\epsfxsize=3in
\epsfbox{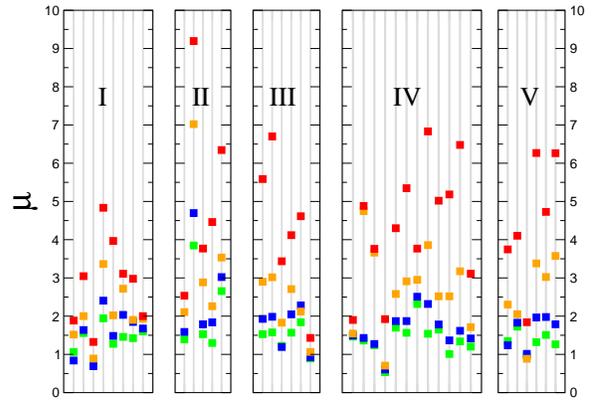}
\vspace{0.3cm}
\caption{Ratio $\mu$ between the percentage of IRs found in non-coding
regions and the percentage of non-coding regions present in the 
entire genome. Each vertical line refers to a genome. 
The 37 investigated genomes are grouped as archaebacteria (I), 
chlamydiae (II), firmicutes (III), proteobacteria (IV) and others (V).
Within each group, the order of the different genomes is 
the same as the one given in Table I.
Different colors of the symbol refer to different 
values of $\ell$. Specifically, we use green 
for $\ell=6$, blue for $\ell=7$, orange  
for $\ell=8$, and red for $\ell>8$.}
\label{fig2}
\end{figure}
The observed excess of the number of IRs 
in most of the complete genomes of eubacteria cannot be
due to a statistical fluctuation. Hence, it is important to relate 
this statistical observation to known biological functions. 
One explanation of the abundance of such 
inverted repeats in eubacteria is that they are used to code
rho-independent hairpin in RNA during protein transcription 
\cite{Carafa90,Schroth95,Cox97,Ermolaeva2000}.
Motivated by this fact, we investigate the location
of each inverted repeats we find with respect to the
biological information present in the annotation of
each  genome. The first investigation concerns the percentage
of inverted repeats located in the non-coding DNA regions
for each genome and for different values of the stem 
length $\ell \ge6$. The results are summarized in Figure 2 
where we show for each genome the 
ratio $\mu$ between the percentage of IRs found in non-coding
regions and the percentage of non-coding regions present in the 
entire genome. Each vertical line refers to a genome. The
grouping of genomes and their sequence is the same as the
one given in Table I. Different colors refer to different 
values of $\ell$. Specifically, we use green squares for $\ell=6$,  
blue squares for $\ell=7$, orange squares for $\ell=8$,
and red squares for $\ell>8$.
 A null random hypothesis would suggest that
$\mu \approx 1$ for all values of $\ell$. Several genomes
of groups II, III, IV and V show values of $\mu$ which are 
much larger than one. The value of $\mu$ always increases
when $\ell$ increases assuming largest values  
for large values of $\ell$.
For most genomes, this implies that 
a large number of inverted repeats are almost
exclusively located in the non-coding regions when the stem length
is longer than 8. However, this behavior is not so pronounced
in all investigated genomes.
It is worth noting that the inverted repeats of 
most archaebacteria are characterized by moderately large
values of $\mu$ (see group I in Figure 2). Moreover, this different
behavior is also observed for some eubacteria such as, for example, 
{\it Mycobacterium tubercolosis}, and {\it Treponema pallidum}. 

A more quantitative test is obtained by performing a $\chi^2$ test
of the null hypothesis that the number on IRs found in non-coding
regions is proportional to the percentage of non-coding regions present
in the entire genome. The $p$-values associated to the above $\chi^2$ test 
for $\ell=6$ are below $1\%$ confidence level in all but three genomes, 
specifically {\it Aeropyrum pernix}, {\it Pseudomonas aeruginosa} 
and {\it Treponema pallidum}. For IRs with stem length $\ell>8$ 
only {\it Halobacterium}
sp. and {\it Treponema pallidum} have a $p$-value larger than $1\%$. This
test shows that for all the value of $\ell$ considered the IRs are
preferentially located in non-coding DNA.

Non-coding regions of complete genomes can be classified with respect to
the orientation of the bounding coding regions. Coding regions can be
``divergent'' (two 5' ends bounding the non-coding region), 
``unidirectional'' (examples are a spacer between two genes
in one operon or a spacer between two consecutive unidirectional
operons) or
``convergent'' (two 3' ends bounding the non-coding region). 
The analysis of the statistical location of IRs in these
biologically different non-coding regions provides relevant
information about the potential biological role of
some of the detected IRs. To perform a statistical analysis 
of the location of IRs in these three types of spacer
we split the set of the non-coding regions in three subsets: 
(i) non-coding regions between two 5' ends of
two ``divergent'' genes (we address these non-coding regions as
type A regions); (ii) non-coding regions between a 3' end and 
a 5' end of two different genes (type B regions) and (iii) 
non-coding regions between two 3' ends of two ``convergent'' genes
(type C regions). 
The genes bounding non-coding regions of type A and C
certainly belong to two different operons, whereas the 
genes bounding non-coding regions of type B may or may not belong
to the same operon. The statistical properties
of the length of the non-coding regions are different for 
the three groups. The non-coding regions
belonging to type A are in average longer than the non-coding
regions belonging to the
other two groups. A statistical characterization of all genomes
shows that in (almost) all
the considered genomes the the probability density function 
of the length of non-coding regions of type A shows a 
broad maximum at a length of 150-200
nucleotides and decays to zero for small and large values of 
non-coding region length. The probability density function of length of
non-coding regions belonging to type B is an exponentially decaying
function approximately. The non-coding regions belonging to type C
behave differently in genomes of
different organisms. In many purple bacteria the probability density function of length of 
non-coding regions has a
sharp peak located at a length of 40-60 nucleotides approximately, whereas in
other genomes an approximately monotonic decaying behavior is observed.

We investigate the distribution of IRs in non-coding regions 
belonging to different groups A, B and C. 
We make the null hypothesis that the probability of having an 
IR in any non-coding part of the genome is just proportional 
to the length of that non-coding region. For example, the probability 
of finding an IR in a non-coding region belonging to type A 
is given by the total length of non-coding regions of this type 
divided by the total length of the non-coding regions of the genome. 
For each genome and for each value of $\ell$ ($\ell=6$,
$\ell=7$, $\ell=8$ and $\ell>8$) we compute the $p$-value associated to the
$\chi^2$ test of the above null hypothesis. Table II shows 
these $p$-values for the considered genomes. 
We note that in archaebacteria the IRs are distributed
in the three groups of non-coding regions in good agreement
with the null hypothesis. The only exception is the {\it Halobacterium}
genome for $\ell=6$ and $\ell>8$. On the
other hand, for many eubacteria the $p$-values are very small especially 
for longer IRs. In these genomes, a direct inspection of the number 
of inverted repeats 
located in each group of non-coding regions shows that the
IRs tend to be preferentially located in non-coding regions of type C 
when $\ell > 8$ for the genomes disproving the null hypothesis. 
Low $p$-values are also observed for low values of $\ell$
(we compute $p$-values also for $\ell=4$ and 5. They are not shown 
in Table II for lack of space but available on-line).
The reason of these low $p$-values for low values of $\ell$ 
is different from the one for the large values of
$\ell$ for most genomes. For example, most of the genomes
(31 over 37) show that the number of IRs located in the type A
non-coding regions is exceeding the number expected under the 
null hypothesis by a six percent in average for $\ell=4$.
This result suggests a potential biological role of short IRs
located in type A non-coding regions which is different from 
the biological role of long IRs located in type C non-coding 
regions.  

Next we investigate the statistical properties of the distance 
of the IRs located in non-coding regions from the 
nearest coding region. This investigation is
performed by dividing IRs in 4 groups.
The first two groups contain IRs located in type A and type C 
non-coding regions
previously defined. IRs located in type A non-coding regions 
are denoted as A5' whereas IRs located in type C non-coding
regions are denoted C3'.
IRs found in type B regions are divided in two subgroups 
depending on the condition that the considered IRs
is closer to a 5' end (we address this subset as B5')
or to a 3' end (we address this subset as B3').

For each group and for each value of stem length $\ell$, 
we estimate the mean distance of IR from the closest
coding regions. This is done by analysing the four groups of 
IRs separately. To obtain mean values which 
are statistically reliable for IRs characterized by both small and
large values of $\ell$ , we perform our analysis on the six genomes
having the largest number of inverted repeats. These genomes 
are {\it Bacillus halodurans}, {\it Bacillus subtilis}, 
{\it Neisseria meningitidis} serogroup B, 
{\it Escherichia coli}, {\it Pseudomonas aeruginosa} and 
{\it Vibrio cholerae} Chr I. In Table III we summarize the average distance
of the IRs from the closest gene boundary for the four groups 
described above for $\ell=6$ and $\ell > 8$. We select these values
of $\ell$ to limit the size of the Table and because they are 
representative of the behavior observed for small and large values of 
$\ell$. From the Table we note that the average distance of
the IRs from the nearest coding region decreases when two conditions
are simultaneously fulfilled. The first is that the considered IR
has a stem length longer than 8 nucleotides ($\ell>8$) and the
second is that the IR is nearby the 3' end of a coding regions. 
In fact, for the B3' subset we observe that the mean distance 
from the 3' end of the coding region decreases when $\ell$ 
increases from 6 to a value larger than 8 for all the 
considered genomes. 
The same behavior is observed in the C3' subset. Indeed in this
last case the decrease of the mean distance is even more
pronounced than in the previous one. 
On the contrary, when the IR is closer to a 5' end of the nearest
coding region the mean distance 
remains approximately the same when $\ell$ increases from 6 to 
values larger than 8 both for the case A5' and for the case B5'.
Only one exception to this general trend is detected. It is the
case of IRs of {\it V. cholerae} Chr I in the B5' subset. 
\begin{figure}[t]
\epsfxsize=3in
\epsfbox{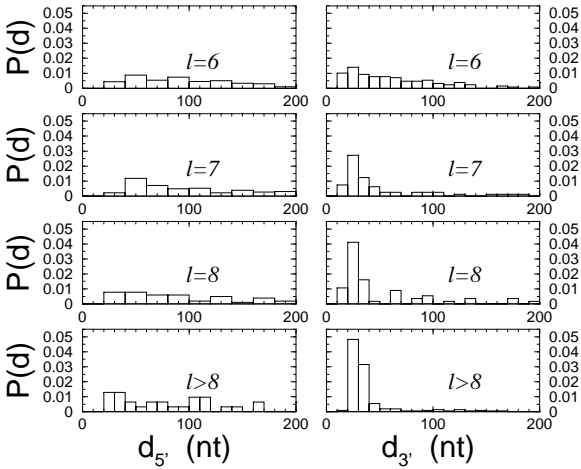}
\vspace{0.3cm}
\caption{Probability density functions $P(d)$ to find
 an IR nearby the 5' end 
of a coding region in a type A non-coding region 
(left panels) or an IR nearby the 3' end 
of a coding region in a type C non-coding region 
(right panels) when 
$\ell=6$, $\ell=7$, $\ell=8$, and $\ell>8$. Data 
refers to the complete genome of {\it E. coli}.}
\label{fig3}
\end{figure}
In Figure 3 we illustrate 
the decrease of the mean distance occurring for most 
eubacterial genomes rich of long inverted repeats 
by considering the case of
{\it E. coli}. In the figure we show the empirical 
probability density functions  $P(d)$ of (i) the 
distance $d_{5'}$ from a 5' end of an IR located
in a type A non-coding region (left panels) and 
(ii) the distance $d_{3'}$ from a 3' end of an IR located
in a type C non-coding region (right panels)
for $\ell=6$, $\ell=7$, $\ell=8$, and $\ell>8$.
The left panels are always broad probability density functions not characterized
by a sharp distance. In the right panels, the probability density function is also 
rather broad for $\ell=6$. However, when $\ell$ increases 
above six $P(d_{3'})$ 
progressively displays a clear peak localized around 
$d_{3'} \approx 20$ nt.
The preferential localization of the IRs
characterized by a long stem nearby the 3' end of a coding
regions supports the biologically motivated hypothesis 
that these structure may play the role of intrinsic terminators
of the transcription process. However, the parallel analysis of 
37 complete genomes summarized in Tables I and II 
shows that this is not a general feature 
of all bacteria but rather depends on the specific
species investigated.

\begin{figure}[t]
\epsfxsize=3in
\epsfbox{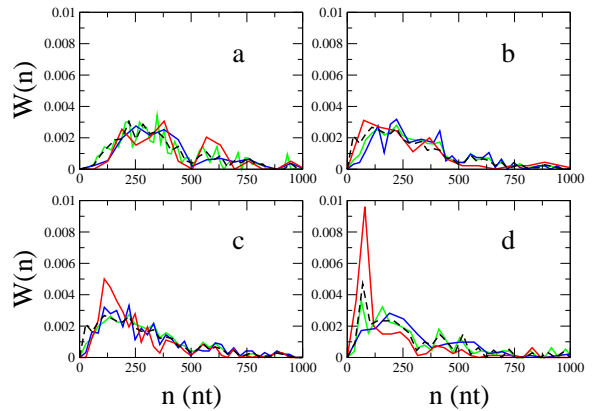}
\vspace{0.3cm}
\caption{
Probability density function $W(n)$ of the length of the non-coding
regions $n$ in which 
an inverted repeat of stem length $\ell=4$ (green line), $\ell=6$ (blu line)
and $\ell>8$ (red line) is found. Data refers to the genome 
of {\it E. coli}. 
The black dashed line is obtained from the null random hypothesis
discussed in the text.  
The four panels show the data for the four subsets of IRs defined in the
text. Specifically, A5' IRs are in (a), B5' in (b), B3' in (c) 
and C3' in (d).}
\label{fig4}
\end{figure}
We have shown that for several eubacteria the average distance 
from the nearest coding region
of IRs characterized by a large value 
of the stem length (usually $\ell>8$)
decreases when the IR is located nearby a 3' end of a coding 
regions. This behavior may be
due to two different hypothesis. First, the longest IRs locate 
preferentially in short non-coding regions. 
Second, for each non-coding region of a given length
the longest IRs belonging to subsets B3' and C3' 
tend to be located closer to the 3' end of the
bounding gene than expected under a random assumption. 
The test of the second hypothesis is statistically difficult because the
number of IRs is not sufficient to give a robust estimation of the mean
distance of the IRs from the 3' end of the bounding gene conditioned to
the length of the non-coding region in which the IR is found. For this
reason, we test only the first hypothesis in the present study.  
In order to test the first hypothesis we investigate the probability $W(n)$ 
that an IR, belonging to one of the four groups previously defined,
is found in a non-coding region of length $n$. Specifically,
for each IR found in non-coding DNA we associate
the length of the non-coding region in which the IR is located. 
We determine the
histogram of the length of the non-coding regions corresponding
to each IR for the stem length $\ell=4$, $\ell=6$ and $\ell>8$ 
for the four subsets A5', B5', B3' and C3'.
This is shown in Figure 4 for the genome 
of {\it E. coli}. The used values of $\ell$ have been selected because 
they are representative of the behavior observed for small, medium 
and large values of the stem length.
As a null hypothesis we
assume that the probability that an IR, which is belonging to one of
the four subsets, is found in a non-coding region of a length $n$ 
is equal to the length of the non-coding region divided by the 
total length $n_{tot}$ of the non-coding regions of the 
same type (A, B or C).
The Figure 4 shows the probability density function expected according to
this null hypothesis as a black dashed line. 
This reference probability density function 
\begin{equation}
W(n)=A \frac{n}{n_{tot}} P(n)
\end{equation}
is obtained by multiplying the empirical
probability density function $P(n)$ of the length of non-coding regions measured 
in the corresponding group of regions  times $n/n_{tot}$, i.e.
the length of the non-coding region divided by the total 
length of non-coding regions belonging to the considered group. 
The parameter $A$ is a normalizing constant.  
Panels (a) and (b) show 
that the probability density function $W(n)$ of IRs of subsets A5' and B5'
is described quite well by the null hypothesis for all values of $\ell$. 
Panel (c) shows that the probability density function $W(n)$ of IRs of subset B3' is
in good agreement with the 
null hypothesis for $\ell=4$ and $\ell=6$ whereas a small discrepancy 
is observed for $\ell>8$.  This discrepancy becomes extremely 
evident in panel (d) which is showing $W(n)$ for
IRs belonging to subset C3'. From the analysis of Figure 3 
we conclude that long IRs located in a type C non-coding
region are found in short non-coding regions 
with a much higher probability than expected from 
a null random hypothesis.

The final investigation concerns a comparison between the 
IRs found by us in the non-coding regions
of several eubacteria and the intrinsic terminators
predicted with bioinformatics methods in Ref. 
\cite{Ermolaeva2000}. In this study the biological
information summarized in the review paper of Ref. 
\cite{Carafa90} are used to train a computer program
able to detect potential rho-independent intrinsic 
terminators in DNA sequences. The optimization 
of the parameters of the program was performed
to ensure that 89\% of the rho-independent intrinsic
terminators known from biological studies are 
identified by using the 98\% confidence threshold. 
We have used the computer program
TransTerm of Ref. \cite{Ermolaeva2000} to detect how many of 
the IRs with $\ell>8$ we find in the non-coding 
regions of several eubacterial genomes are predicted as 
rho-independent terminators by 
TransTerm.
\begin{figure}[t]
\epsfxsize=3in
\epsfbox{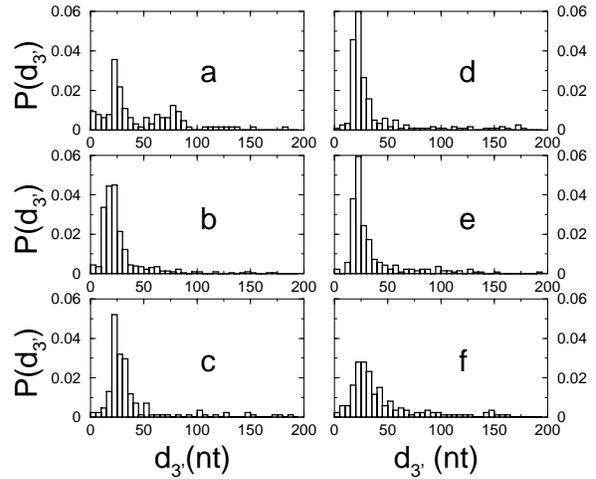}
\vspace{0.3cm}
\caption{Probability density function $P(d_{3'})$ of a subset of
IRs with stem length $\ell>8$ located in type B and C non coding regions. 
The subset is defined by considering the IRs which are nearby the 3' end 
of a coding region and
which are not predicted as rho-independent intrinsic terminators 
by the TransTerm program at a 98\% confidence threshold 
in the six genomes of 
(a) {\it Bacillus halodurans} (129 IRs),
(b) {\it Bacillus subtilis} (459 IRs), 
(c) {\it Neisseria meningitidis} serogroup B (169 IRs), 
(d) {\it Escherichia coli} (241 Irs), 
(e) {\it Pseudomonas aeruginosa} (279 IRs) and 
(f) {\it Vibrio cholerae} (172 IRs).}
\label{fig5}
\end{figure}
By selecting the confidence level of 98\% recommended
by the authors, we note that only a part of the IRs 
localized nearby the end of the nearest coding 
regions are predicted as rho-independent terminators 
by TransTerm.
Hundreds of IRs remain not predicted as 
potential rho-independent terminators by TransTerm.
We verify that these IRs maintain the characteristic of being 
localized at a typical distance from the end of the
nearest coding regions in several genomes very 
rich in IRs. Figure 5 shows the inverted 
repeats with $\ell>8$ which are not predicted as 
rho-independent intrinsic terminators by TransTerm
in the six genomes of {\it Bacillus halodurans},
{\it Bacillus subtilis}, {\it Neisseria meningitidis} serogroup B, 
{\it Escherichia coli}, {\it Pseudomonas aeruginosa} and 
{\it Vibrio cholerae}. In all these cases the IRs are 
still localized near the end of 
the nearest coding regions.

\subsection*{Conclusion}

In summary, a comparative statistical investigation of the number
and location of IRs in several different
complete genomes shows that 
a large number of them may play a role in several
eubacteria as intrinsic terminators of the transcription 
process. 
For almost all the bacteria investigated, we quantitatively 
show that the IRs locates preferentially in non-coding regions.
Moreover, in several eubacteria, IRs characterized by 
large values of the stem length $\ell$ locate preferentially
in the non-coding regions bounded by two 3' ends of convergent 
genes. We also show statistical evidence that long 
IRs are found in short non-coding region more frequently than
expected under a null random hypothesis.   
A large number of the IRs detected with 
our statistical methodology near the 3' end of a coding region
are different from
the rho-independent transcription terminators
detected with a specialized computer program
trained by using molecular biological information 
exclusively determined by investigating
transcription process in {\it Escherichia coli}.
Our findings based on the comparative analysis 
of non-coding regions of complete genomes suggest 
that other forms of intrinsic termination
may be active in several eubacteria.

Inverted repeats are not only involved in
transcription termination in bacteria. Indeed, by
performing our comparative genomic study we are able
to show that short IRs are slightly more abundant 
than expected under
a null hypothesis in type A non-coding regions of 
several genomes. This may be due to nucleotide
concentration fluctuation correlated to the type of
non-coding region considered or may indicate a potential
biological role of some of these short IRs consistent with
known results of the literature 
\cite{Post78,Shen88,Li97,Sadler83,Simons84}. 
The role of IRs with short values of $\ell$ 
located in type A non-coding regions will be 
considered in a future study.

\section*{Acknowledgments}

We thank INFM and MURST for financial support.
We also wish to thank referees for useful remarks
and interesting suggestions.
This work is part of the INFM-PAIS project ``Statistical 
modeling of non-coding DNA''.

\twocolumn[\hsize\textwidth\columnwidth\hsize\csname@twocolumnfalse\endcsname

 \begin{table}
\caption{Inverted repeats per million base
pair of stem length $\ell$ and loop length  within the
interval $3\le m \le10$ detected in 37 complete genomes.}
\begin{tabular}{l|rrrr}
 Genome&$l=6$&$l=7$&$l=8$&$l>8$\\
 \tableline
{\it A. pernix} &\textcolor{rosso}{ 1519.44~(1388.88)} &\textcolor{rosso}{  444.39~( 344.97)}&\textcolor{black}{  116.19~(110.20)} &\textcolor{rosso}{   45.52~( 26.95)}\\
{\it A. fulgidus} &\textcolor{black}{ 1387.72~(1325.74)} &\textcolor{black}{  302.52~( 331.89)}&\textcolor{black}{   94.11~( 81.71)} &\textcolor{black}{   25.71~( 34.89)}\\
{\it Halobacterium} sp. &\textcolor{rosso}{ 2893.40~(2450.06)} &\textcolor{rosso}{  938.32~( 728.31)} &\textcolor{rosso}{  328.66~(215.96)} &\textcolor{rosso}{  171.78~( 86.88)}\\
{\it M. jannaschii} &\textcolor{black}{ 2258.90~(2378.42)} &\textcolor{black}{  706.32~(  689.50)}&\textcolor{black}{  219.82~(211.42)} &\textcolor{rosso}{  137.54~( 81.08)}\\
{\it M. thermoautotrophicum~} &\textcolor{rosso}{1573.62~( 1272.71)} &\textcolor{rosso}{  435.66~( 341.45)} &\textcolor{rosso}{  108.49~( 80.51)} &\textcolor{black}{   34.83~( 23.98)}\\
{\it P. abyssi} &\textcolor{black}{ 1291.13~(1328.52)} &\textcolor{black}{  339.35~( 328.59)}&\textcolor{black}{   92.91~( 83.28)} &\textcolor{black}{   39.66~( 31.16)}\\
{\it P. horikoshii} &\textcolor{black}{ 1396.03~(1414.43)} &\textcolor{black}{  371.58~(  389.99)}&\textcolor{black}{  101.81~(104.69)} &\textcolor{black}{   48.32~( 40.84)}\\
{\it T. acidophilum} &\textcolor{rosso}{ 1455.68~(1229.47)} &\textcolor{rosso}{  398.11~(  318.87)}&\textcolor{rosso}{  114.38~( 78.60)} &\textcolor{black}{   46.65~( 34.51)}\\
 \tableline
{\it C. pneumoniae} AR39 &\textcolor{viola}{ 1916.49~(1428.63)} &\textcolor{rosso}{  541.53~(  366.71)} &\textcolor{rosso}{  187.83~(   94.32)} &\textcolor{viola}{  135.79~( 31.71)}\\
{\it C. pneumoniae} CWL029 &\textcolor{viola}{ 1915.90~(1402.99)} &\textcolor{rosso}{  541.36~(  403.99)} &\textcolor{rosso}{  188.58~(101.61)} &\textcolor{viola}{  132.50~( 24.39)}\\
{\it C. pneumoniae} J138 &\textcolor{viola}{ 1921.41~(1465.48)} &\textcolor{rosso}{  538.16~(  381.84)} &\textcolor{viola}{  188.88~(   86.30)} &\textcolor{viola}{  135.15~( 33.38)}\\
{\it C. trachomatis} &\textcolor{viola}{ 1997.36~(1439.11)} &\textcolor{rosso}{  575.08~(  401.16)} &\textcolor{viola}{  205.72~(   98.18)} &\textcolor{viola}{  165.51~( 29.92)}\\
{\it C. trachomatis} ser D &\textcolor{viola}{ 1956.80~(1391.82)} &\textcolor{viola}{  638.84~(  335.73)} &\textcolor{viola}{  195.68~(   95.92)} &\textcolor{viola}{  175.54~( 34.53)}\\
 \tableline
{\it B. halodurans} &\textcolor{rosso}{ 1415.16~(1289.04)} &\textcolor{black}{  376.69~(  340.29)} &\textcolor{black}{  103.04~(   90.90)} &\textcolor{viola}{  143.02~( 32.36)}\\
{\it B. subtilis} &\textcolor{black}{ 1492.83~(1408.13)} &\textcolor{black}{  404.76~(  355.41)} &\textcolor{black}{  121.95~(101.07)} &\textcolor{viola}{  216.14~( 28.47)}\\
{\it M. genitalium} &\textcolor{viola}{ 3192.70~(2628.98)} &\textcolor{viola}{ 1030.90~(  656.81)} &\textcolor{rosso}{  330.99~(237.90)} &\textcolor{rosso}{  160.32~( 89.64)}\\
{\it M. pneumoniae} &\textcolor{viola}{ 2163.17~(1618.09)} &\textcolor{rosso}{  608.77~(  436.06)} &\textcolor{viola}{  224.16~(104.12)} &\textcolor{rosso}{  107.79~( 53.90)}\\
{\it U. urealyticum} &\textcolor{rosso}{ 4244.94~(3824.57)} &\textcolor{rosso}{ 1416.75~(1219.87)} &\textcolor{rosso}{  525.46~(401.75)} &\textcolor{viola}{  323.26~(183.58)}\\
{\it M. tubercolosis} &\textcolor{rosso}{ 2179.97~(1947.85)} &\textcolor{rosso}{  638.55~(  560.80)} &\textcolor{rosso}{  202.20~(158.45)} &\textcolor{rosso}{  100.87~( 56.67)}\\
 \tableline
{\it R. prowazekii} &\textcolor{rosso}{ 3131.74~(2724.19)} &\textcolor{rosso}{  957.25~(  853.78)} &\textcolor{rosso}{  287.89~(229.41)} &\textcolor{rosso}{  166.44~(103.46)}\\
{\it N. meningitidis} ser A &\textcolor{rosso}{ 1649.88~(1463.56)} &\textcolor{black}{  432.61~(  397.36)} &\textcolor{viola}{417.96~(   96.14)} &\textcolor{viola}{  231.64~( 39.37)}\\
{\it N. meningitidis} ser B &\textcolor{rosso}{ 1648.07~(1419.68)} &\textcolor{rosso}{  462.52~(  375.38)} &\textcolor{viola}{415.43~(   94.62)} &\textcolor{viola}{  239.40~( 34.77)}\\
{\it C. jejuni} &\textcolor{rosso}{ 3334.18~(2939.42)} &\textcolor{rosso}{ 1019.81~(  883.35)} &\textcolor{rosso}{  339.33~(252.82)} &\textcolor{rosso}{  226.62~(122.45)}\\
{\it H. pylori} 26695 &\textcolor{rosso}{ 2159.04~(1960.59)} &\textcolor{black}{  585.78~(  549.80)} &\textcolor{black}{  204.45~(175.07)} &\textcolor{rosso}{   90.53~( 65.35)}\\
{\it H. pylory} J99 &\textcolor{rosso}{ 2101.19~(1885.84)} &\textcolor{black}{  605.29~(  537.16)} &\textcolor{rosso}{  184.33~(147.22)} &\textcolor{rosso}{   83.34~( 55.97)}\\
{\it Buchnera} sp. &\textcolor{viola}{ 4807.38~(3469.75)} &\textcolor{viola}{ 1793.40~(1052.01)} &\textcolor{viola}{  616.53~(334.02)} &\textcolor{viola}{  358.99~(163.89)}\\
{\it E. coli} &\textcolor{rosso}{ 1441.62~(1216.58)} &\textcolor{rosso}{  419.90~(  322.04)} &\textcolor{rosso}{  147.22~( 76.52)} &\textcolor{viola}{  154.98~( 27.16)}\\
{\it H. influenzae} &\textcolor{rosso}{ 2015.69~(1748.50)} &\textcolor{rosso}{  596.13~(  469.36)} &\textcolor{rosso}{  181.95~(139.88)} &\textcolor{viola}{  204.36~( 62.84)}\\
{\it P. aeruginosa} &\textcolor{rosso}{ 2422.58~(2072.34)} &\textcolor{rosso}{  740.37~(  563.02)} &\textcolor{rosso}{  251.74~(167.29)} &\textcolor{viola}{  182.30~( 58.90)}\\
{\it V. cholerae} Chr I &\textcolor{rosso}{ 1508.87~(1266.40)} &\textcolor{rosso}{  414.37~( 323.19)} &\textcolor{rosso}{  136.77~( 85.10)} &\textcolor{viola}{  215.79~( 29.04)}\\
{\it X. fastidiosa} &\textcolor{rosso}{ 1391.03~(1155.52)} &\textcolor{rosso}{  397.12~(  301.20)} &\textcolor{rosso}{  103.01~( 74.65)} &\textcolor{rosso}{   76.51~( 27.99)}\\
 \tableline
{\it A. aeolicus} &\textcolor{black}{ 1489.68~(1518.69)} &\textcolor{black}{  415.77~(  384.83)} &\textcolor{black}{  107.00~(109.58)} &\textcolor{black}{   35.45~( 35.45)}\\
{\it B. burgdorferi} &\textcolor{rosso}{ 3235.89~(2847.19)} &\textcolor{rosso}{ 1068.38~(  908.07)} &\textcolor{black}{  332.70~(311.84)} &\textcolor{viola}{  262.43~(124.08)}\\
{\it T. pallidum} &\textcolor{rosso}{ 1539.53~(1213.52)} &\textcolor{rosso}{  424.42~(  304.04)} &\textcolor{black}{  100.17~( 76.45)} &\textcolor{viola}{   69.42~( 21.09)}\\
{\it D. radiodurans} &\textcolor{bluuu}{ 1545.32~(2050.11)} &\textcolor{bluuu}{  345.46~(  584.83)} &\textcolor{bluuu}{   81.93~(161.59)} &\textcolor{black}{   62.30~( 65.32)}\\
{\it Synechocystis} sp. &\textcolor{bluuu}{ 1096.13~(1379.61)} &\textcolor{bluuu}{  223.59~(  378.34)} &\textcolor{bluuu}{   47.01~( 94.59)} &\textcolor{bluuu}{   12.03~( 36.94)}\\
{\it T. maritima} &\textcolor{rosso}{ 1609.05~(1399.99)} &\textcolor{rosso}{  447.14~(  382.65)} &\textcolor{rosso}{  140.81~(107.48)} &\textcolor{viola}{  116.62~( 25.26)}\\
 \end{tabular}
 \end{table}
 The number in parenthesis is the theoretical prediction
obtained from computer data generated by using a local
first-order Markov process. We use the color code described 
in the text to quantify 
the discrepancy between the empirical results
and the predictions of the Markov model.
\vskip2pc]

\newpage

\twocolumn[\hsize\textwidth\columnwidth\hsize\csname@twocolumnfalse\endcsname
 \begin{table}
 \caption{$p$-values associated with 
the $\chi^2$ test of the
null hypothesis discussed in the text on the distribution of 
inverted repeats in the different
types of non-coding regions for 37 bacterial genomes.}
 \begin{tabular}{l|rrrr}
 $Genome$ &$l=6$&$l=7$&$l=8$&$l>8$\\
 \tableline
{\it A. pernix}  &     69.68 &     49.54 &     78.37 &     82.65  \\
{\it A. fulgidus}  &     25.53 &     44.92 &      6.75 &     75.39 \\
{\it Halobacterium}  &      0.47 &     19.94 &     61.33 &      0.33  \\
{\it M. jannaschii}  &     42.14 &      2.33 &      3.09 &     90.29  \\
{\it M. thermoautotrophicum~} &      2.16 &     28.95 &      2.66 &     88.31  \\
{\it P. abyssi}  &     54.75 &      7.50 &      3.23 &     41.93  \\
{\it P. horikoshii}  &      6.59 &     23.62 &     70.80 &     39.27  \\
{\it T. acidophilum} &      5.80 &      2.50 &     27.14 &     69.38  \\
\tableline
{\it C. pneumoniae} AR39  &     10.70 &     37.87 &      6.16 &      8.28  \\
{\it C. pneumoniae} CWL029 &      1.38 &     12.45 &     11.98 &      4.35  \\
{\it C. pneumoniae} J138  &      7.01 &     29.77 &     42.17 &      0.00  \\
{\it C. trachomatis}  &     70.56 &     79.23 &     89.88 &      0.00  \\
{\it C. trachomatis} ser D &      0.85 &     48.25 &     95.20 &      0.00  \\
\tableline
{\it B. halodurans} &     10.10 &     50.04 &     88.75 &      0.00  \\
{\it B. subtilis}  &      8.03 &     49.84 &      1.72 &      0.00  \\
{\it M. genitalium} &     97.27 &     23.22 &     22.19 &     24.69  \\
{\it M. pneumoniae} &     11.97 &     60.14 &     38.18 &     28.93  \\
{\it U. urealyticum} &      0.18 &     21.55 &      9.51 &      0.48  \\
{\it M. tubercolosis}&     86.16 &     23.36 &     70.71 &      3.73  \\
\tableline
{\it R. prowazekii}         &      2.19 &     58.18 &     88.54 &     87.07  \\
{\it N. meningitidis} ser A &     13.71 &     21.99 &      0.00 &      0.00  \\
{\it N. meningitidis} ser B &      0.33 &      0.08 &      0.00 &      0.00  \\
{\it C. jejuni}             &      0.02 &     49.08 &     93.59 &      0.00  \\
{\it H. pylori} 26695       &     17.32 &      0.04 &     97.81 &     11.29  \\
{\it H. pylory} J99         &     28.40 &     87.26 &     73.48 &      0.01  \\
{\it Buchnera} sp.          &      0.03 &      4.37 &     42.02 &     11.16  \\
{\it E. coli}               &      3.06 &      0.16 &      0.00 &      0.00  \\
{\it H. influenzae}         &      0.86 &     38.33 &      0.20 &      0.00  \\
{\it P. aeruginosa}         &      1.54 &      0.00 &      0.00 &      0.00  \\
{\it V. cholerae} Chr I     &      0.64 &     19.41 &      0.00 &      0.00  \\
{\it X. fastidiosa} &     78.58 &     35.77 &     17.10 &      2.02  \\
\tableline
{\it A. aeolicus}     &     69.94 &     32.28 &     24.83 &     66.97  \\
{\it B. burgdorferi}  &      0.01 &     14.34 &     25.77 &      0.00  \\
{\it T. pallidum}     &     21.35 &     91.15 &      0.99 &      6.45  \\
{\it D. radiodurans}  &      0.03 &      0.00 &      0.03 &      0.00  \\
{\it Synechocystis} sp.  &      1.46 &     40.93 &      1.59 &     33.15  \\
{\it T. maritima}        &     20.98 &     27.83 &      2.59 &      0.00  \\
 \end{tabular}
\end{table}
\begin{table}
\caption{Mean distance of the inverted repeats from the
closest gene boundary for the four groups of inverted repeats defined in
the text for six bacterial genomes.} 
\begin{tabular}{l|cc|cc|cc|cc}
 \multicolumn{1}{c|}{~}&\multicolumn{2}{c|}{A5'}&\multicolumn{2}{c|}{B5'}&\multicolumn{2}{c|}{B3'}&\multicolumn{2}{c}{C3'}\\
 $Genome$ & $\ell=6$ & $\ell>8~$ & $\ell=6$ & $\ell>8~$ & $\ell=6$ & $\ell>8~$ & $\ell=6$ & $\ell>8~$\\
 \hline
 {\it B. halodurans} & 97.41 &  111.2~ &  106.3 &  106.1~ & 84.49 & 56.28~ & 67.34 & 30.59~\\
 {\it B. subtilis} & 86.84 &  81.90~ &  78.82 &  83.79~ & 61.96 & 30.12~ & 79.75 & 25.86~\\
 {\it E. coli} & 99.71 &  89.64~ &  85.96 &  80.00~ & 74.41 & 39.31~ & 80.85 & 32.19~\\
 {\it N. meningitidis} ser B & 100.9 &  83.19~ &  123.5 &  100.7~ & 97.46 & 41.47~ & 154.0 & 48.92~\\
 {\it P. aeruginosa} & 91.26 &  117.0~ &	 76.01 &  77.64~ & 66.04 & 38.87~ & 80.46 & 35.95~\\
 {\it V. cholerae} Chr I & 93.39 &  110.7~ &  95.52 &  40.96~ & 62.61 & 38.90~ & 84.62 & 41.86~\\
 \end{tabular}
\end{table}
\vskip2pc]

\end{document}